\documentclass[conference]{IEEEtran}
\IEEEoverridecommandlockouts
\usepackage{cite}
\usepackage{amsmath,amssymb,amsfonts}
\usepackage{graphicx}
\usepackage{textcomp}
\usepackage{xcolor}
\usepackage{algorithm}
\usepackage{algpseudocode}
\newcommand{\ignore}[1]{}
\def\BibTeX{{\rm B\kern-.05em{\sc i\kern-.025em b}\kern-.08em
    T\kern-.1667em\lower.7ex\hbox{E}\kern-.125emX}}
    \newcommand{\Name}{SECite}
    
\begin{document}

% \IEEEpubid{\makebox[2\columnwidth]{%
% 979-8-3315-9397-1/26/\$31.00~\copyright~2026 IEEE
% \hfill}\hspace{\columnsep}}

% 979-8-3315-9397-1/26/$31.00 ©2026 IEEE

\title{SECite: Analyzing and Summarizing Citations in Software Engineering Literature\\}

%\ignore{
\author{\IEEEauthorblockN{Shireesh Reddy Pyreddy, Khaja Valli Pathan}
\IEEEauthorblockA{\textit{Dept. of Computer Science} \\
\textit{SUNY Polytechnic Institute}\\
Utica, NY, USA \\
pyredds@sunypoly.edu, pathank@sunypoly.edu}
\and
\IEEEauthorblockN{Hasan Masum, Tarannum Shaila Zaman}
\IEEEauthorblockA{\textit{Dept. of Information Systems} \\
\textit{University of Maryland Baltimore County}\\
Maryland, USA \\
hmasum1@umbc.edu, zamant@umbc.edu}}

\newcommand{\commenttsz}[1]{{\color{blue} \sf (TSZ: #1)}}
\newcommand{\commenthm}[1]{{\color{red} \sf (HM: #1)}}
\newcommand{\writehm}[1]{{\color{red} #1}}
\newcommand{\reviewtwo}[1]{{\color{purple} R2: #1}}

\maketitle
% \IEEEpubidadjcol 

\begin{abstract}
Identifying the strengths and limitations of a research paper is a core component of any literature review. However, traditional summaries reflect only the authors’ self-presented perspective. Analyzing how other researchers discuss and cite the paper can offer a deeper, more practical understanding of its contributions and shortcomings. In this research, we  introduce \Name{}, a novel approach for evaluating scholarly impact through sentiment analysis of citation contexts. We develop a semi-automated pipeline to extract citations referencing nine research papers and apply advanced natural language processing (NLP) techniques with unsupervised machine learning to classify these citation statements as positive or negative. Beyond sentiment classification, we use generative AI to produce sentiment-specific summaries that capture the strengths and limitations of each target paper, derived both from clustered citation groups and from the full text. Our findings reveal meaningful patterns in how the academic community perceives these works, highlighting areas of alignment and divergence between external citation feedback and the authors’ own presentation. By integrating citation sentiment analysis with LLM-based summarization, this study provides a comprehensive framework for assessing scholarly contributions.
%and offers a richer lens for understanding how research shapes, and is shaped by, academic discourse.
\end{abstract}

\begin{IEEEkeywords}
Sentiment Analysis, LLMs, Text Summarization, Citations 
\end{IEEEkeywords}

\section{Introduction}

Feedback is the cornerstone of intellectual progress, driving the exchange of ideas and fostering meaningful scholarly discourse \cite{malcom2018s,staines2018community}. In academic research, understanding how one’s work is perceived within the larger community is essential not only for validation but also for shaping future contributions. However, traditional methods of gathering and analyzing feedback are often constrained by scale and subjectivity. To address these limitations, our study introduces a semi-automated, data-driven framework, \Name{} that leverages unsupervised sentiment analysis and LLM-driven summarization to evaluate citations referencing scholarly work. By uncovering patterns, sentiments, and opinions embedded in these references, our aim is to provide a deeper and more nuanced understanding of academic impact and engagement \cite{yu2013automated,ghosh2017determiningsentimentcitationtext}. 

Our approach focuses on sentiment analysis of citations from research papers related to software engineering \cite{yu2017descry}, revealing how the academic community perceives its contributions. Using Generative AI \cite{banh2023generative}, we go further to generate sentiment-specific summaries, highlighting the positive and negative aspects of the referenced work. This dual-layered analysis not only evaluates the sentiment embedded in individual citations but also offers a holistic view of how the work resonates within its field \cite{kunnath2021meta,10.1007/s11192-022-04567-4,ikram2019aspect}. By integrating these insights, we enable researchers to understand the broader narrative surrounding their work and its influence on academic discourse \cite{parthasarathy2014sentiment,xu2015citation}. 

%\commenttsz{One paragraph comparing with existing works}

%\writehm{
Existing approaches to citation sentiment analysis \cite{yu2013automated, athar2012context, ikram2019aspect, zhou2022research, karim2022comprehension, muppidi2020approach, xu2015citation, parthasarathy2014sentiment} make substantial contributions but primarily target isolated components of the analysis pipeline. Early biomedical studies introduce core sentiment classification techniques \cite{yu2013automated}, while subsequent work strengthens contextual understanding through linguistic patterns and aspect-based methods \cite{athar2012context, ikram2019aspect}. More recent deep learning approaches advance sentiment detection and support domain-specific applications across clinical and large-scale citation datasets \cite{zhou2022research, karim2022comprehension, muppidi2020approach, xu2015citation, parthasarathy2014sentiment}. However, these methods typically treat sentiment classification and summarization as separate tasks. \Name{} addresses this limitation by providing a semi-automated framework that integrates advanced NLP techniques, unsupervised machine learning (K-means clustering)~\cite{hartigan1979algorithm}, and Generative AI~\cite{banh2023generative} to jointly classify citation sentiment and generate dual-perspective summaries that capture both positive and negative aspects of referenced works.
%}

In today’s rapidly evolving research landscape—where disciplines intersect and methodologies diverge, producing knowledge alone is not enough. Understanding how peers and collaborators perceive that knowledge is essential for refining and extending scholarly contributions. Our framework offers an innovative and scalable solution by transforming citation analysis into a powerful mechanism for evaluating and summarizing the reception of academic work \cite{umer2021scientific, nazir2022important}. By aligning sentiment analysis with summarization, this study provides a more comprehensive view of how research influences and shapes the academic community, ultimately supporting the ongoing dialogue that drives intellectual progress \cite{xu2015citation, tandon2012citation}.

%\commenttsz{Summary of the method and result highlights}

%\writehm{
\Name{} delivers citation sentiment analysis with intelligent summary generation through three major steps: (1) semi-automated extraction of citation statements from research papers, (2) NLP-based sentiment classification of citations into positive, or negative using an unsupervised K-means clustering algorithm, and (3) Generative AI–powered summarization that highlights both the strengths and weaknesses of the referenced works.
%}
\section{Background}
% \reviewtwo{Refine language in some sections: Minor grammar issues, long sentences, and repeated explanations (especially in Background).}

\ignore{
\begin{figure}[htbp]
\centerline{\includegraphics[width=0.5\textwidth]{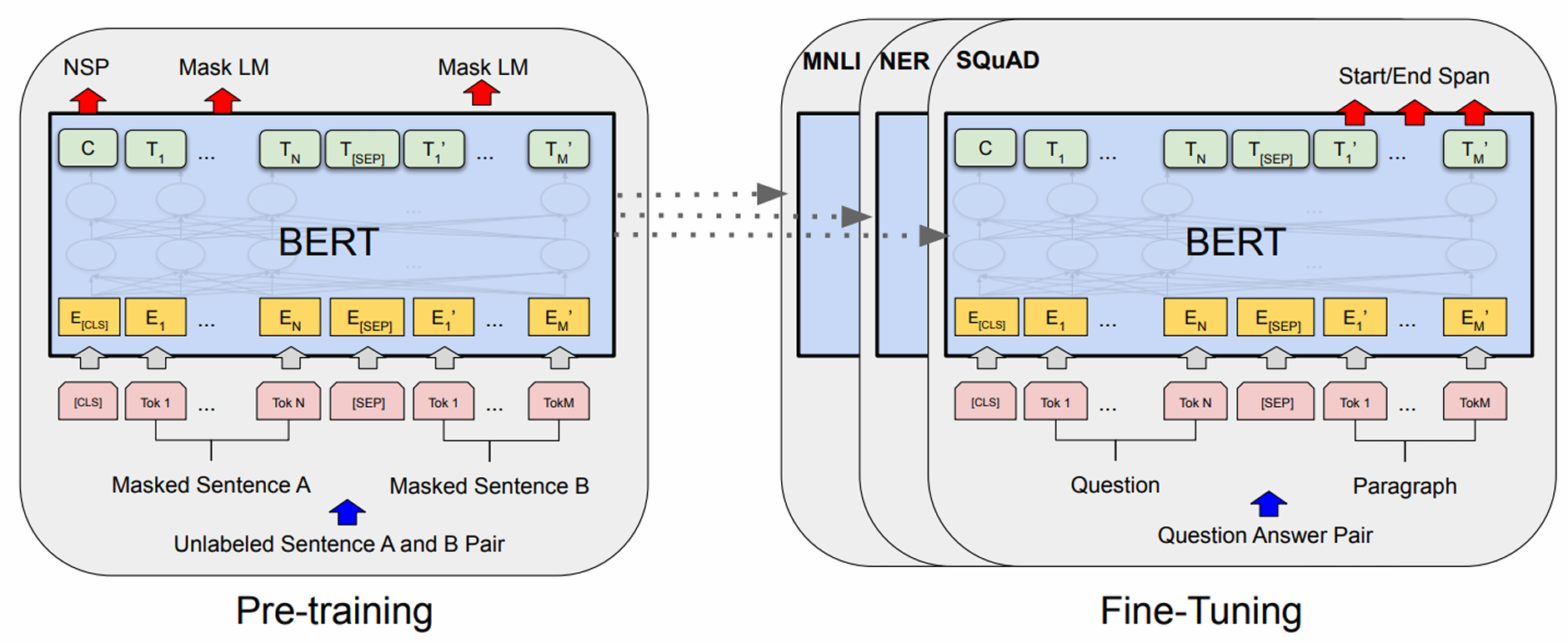}}
\caption{BERT Architecture}
\label{fig}
\end{figure}}
\noindent
\textbf{BERT:} BERT (Bidirectional Encoder Representations from Transformers) \cite{devlin2019bert} is a groundbreaking NLP model developed by Google, known for its ability to understand the context of words by considering both their left and right contexts simultaneously. Built on the Transformer architecture, BERT employs the encoder stack and self-attention mechanisms to process input text efficiently. During pretraining, it uses two main objectives: Masked Language Modeling (MLM), where random words are masked and predicted based on surrounding context, and Next Sentence Prediction (NSP), which helps it understand relationships between sentence pairs. This pretraining equips BERT with a deep understanding of language, which can then be fine-tuned for specific tasks like text classification, question answering, and named entity recognition.
\ignore{
\begin{figure}[htbp]
\centerline{\includegraphics[width=0.5\textwidth]{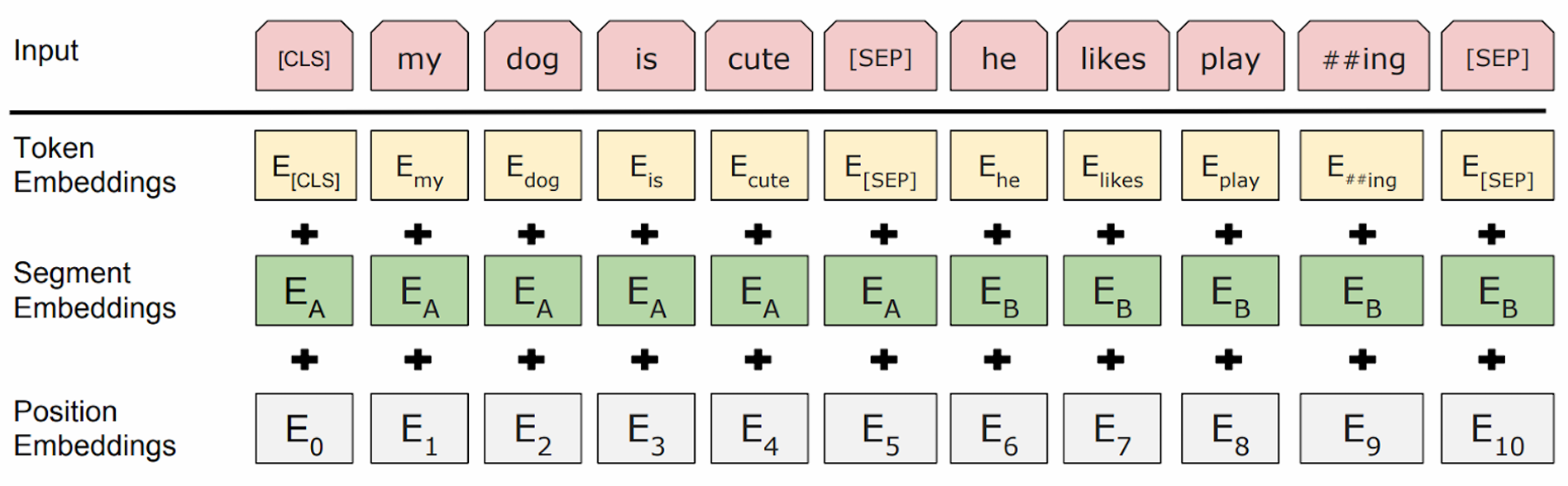}}
\caption{BERT Input Representation}
\label{fig}
\end{figure}
}
BERT's input representation includes token embeddings, segment embeddings, and positional embeddings, with special tokens like [CLS] for classification and [SEP] to separate sentences. Its bidirectional nature and pretraining on massive datasets make it excel in understanding nuanced language tasks. Variants like RoBERTa \cite{liu2019roberta} and DistilBERT \cite{sanh2019distilbert} have optimized BERT further for specific use cases, while its applications range from search engines to chatbots and medical text processing. Despite its strengths, BERT is computationally intensive and challenging to scale for longer inputs, highlighting the trade-off between performance and resource demands \cite{zaheer2020big}.  %\commenthm{cite?}

\noindent
\textbf{K-means:} K-means \cite{hartigan1979algorithm} is a popular clustering algorithm used in machine learning to group data points into k clusters based on their similarity. It works iteratively by initializing k centroids, assigning each data point to the nearest centroid, and then recalculating the centroids based on the mean of the points in each cluster. This process repeats until the centroids stabilize or the changes between iterations fall below a threshold. K-means is particularly effective for segmenting datasets into distinct groups and is commonly used in market segmentation, image compression, and anomaly detection. \cite{ahmed2020k} %\commenthm{cite?}

\ignore{
The k-means clustering algorithm is as follows:

\begin{figure}[htbp]
\centerline{\includegraphics[width=0.5\textwidth]{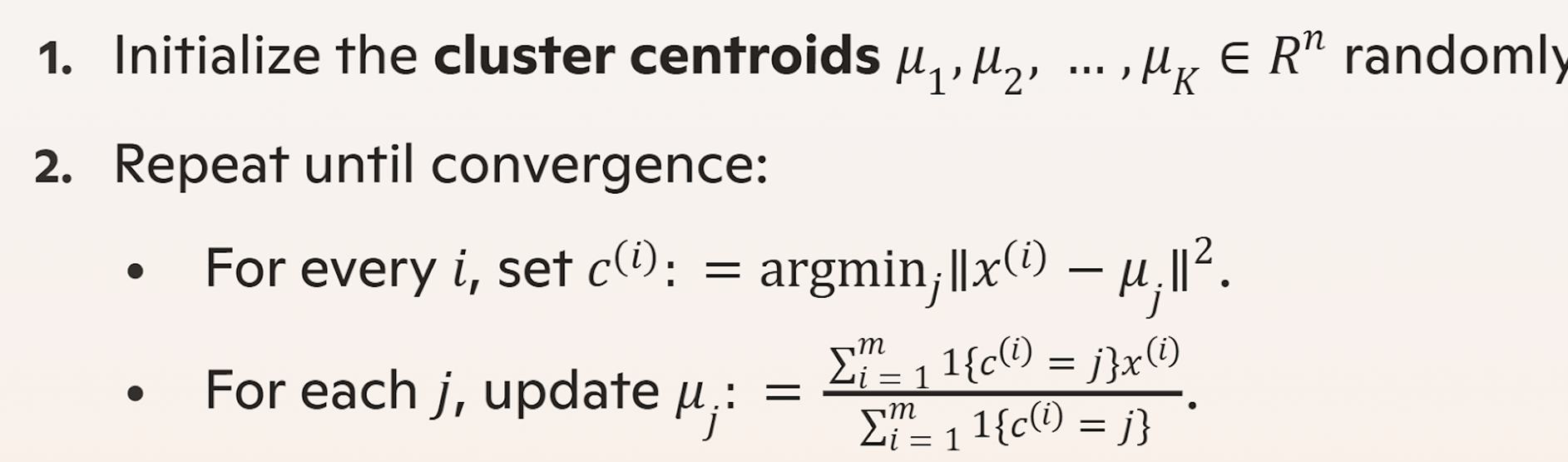}}
\caption{K-Means Algorithm}
\label{fig}
\end{figure}}

%\writehm{removed limitations from here. added in limiation section}
\ignore{
Despite its simplicity and efficiency, K-means has some limitations. The number of clusters, \writehm{k}, must be chosen in advance, and the algorithm is sensitive to the initial placement of centroids, which can lead to suboptimal clustering \cite{arthur2006k}. Additionally, K-means assumes clusters are spherical and equally sized, making it less effective for irregularly shaped or imbalanced datasets \cite{ahmed2020k}. Nonetheless, its speed and ease of implementation make it a fundamental tool for exploratory data analysis and unsupervised learning tasks \cite{arthur2006k}.
}

\noindent
\textbf{t-SNE:} t-SNE (t-Distributed Stochastic Neighbor Embedding) \cite{maaten2008visualizing} is a dimensionality reduction technique designed to visualize high-dimensional data in 2D or 3D. It excels at preserving the local structure of the data, making it particularly effective for clustering and identifying patterns. t-SNE works by first converting the high-dimensional distances between points into probabilities that represent similarities. These similarities are then mapped to a lower-dimensional space, aiming to minimize the divergence between the original and reduced-space distributions using a gradient descent optimization. While t-SNE produces highly interpretable visualizations, it has limitations, including high computational cost for large datasets and sensitivity to hyperparameters like perplexity. Despite this limitation, it is widely used in fields like bioinformatics and natural language processing to explore data structures.

\section{Approach}

\begin{figure}[htbp]
\centerline{\includegraphics[scale=0.54]{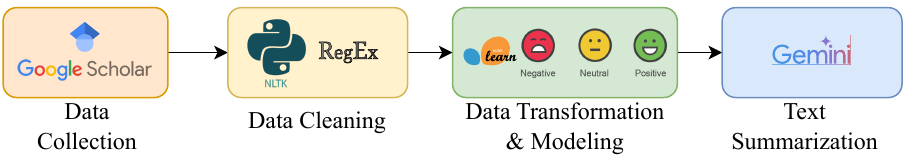}}
\caption{The Overview of \Name}
\label{fig:overview}
\end{figure}

\Name{} gathers and analyzes feedback on a research paper by examining the sentiment\cite{10903889} and tone of citations referenced in other academic works. Figure~\ref{fig:overview} presents the overall workflow of \Name{}, which consists of four major steps: data collection, data cleaning, sentiment analysis, and summary generation using large language models (LLMs). For this analysis, we collect research publications focused on software engineering and debugging\cite{lota5060080recent}.

\subsection{Data Collection}\label{AA}

%The initial phase of the process involves data collection, achieved by designing a semi-automated pipeline using Python, PyPDF and Selenium to extract and parse papers from Google Scholar. This stage presented several significant challenges presented in detail below. Each of these challenges demanded meticulous attention and innovative solutions to ensure data integrity and comprehensiveness. Ultimately, we succeeded in automating approximately 70\% of the process, with the remaining 30\% being completed manually. 

We collect 911 citation references from 151 papers and design a semi-automated pipeline that interfaces with Google Scholar to efficiently extract citation information. The procedure consists of several precise steps to ensure comprehensive data acquisition:\\
\textbf{Automated Navigation and Search:} The pipeline opens the Google Scholar website using Selenium for browser automation. It enters the target research paper title and searches to locate the exact paper.\\
\textbf{Accessing Citation References:} After identifying the correct paper, the system clicks on the citations link to access all citing works. This step is crucial for collecting a comprehensive set of citation sources.\\
\textbf{Link Collection and Paper Retrieval:} The pipeline gathers the links to all citing papers, sequentially opens each link, downloads the associated paper, and saves it to a designated location for processing.\\
\textbf{PDF Parsing and Extraction:} Each downloaded paper is loaded and processed through a custom PDF parser function implemented in Python. This function extracts citations, ensuring that relevant text and references are identified and stored for analysis.\\
\textbf{File Management:} After extraction, the system deletes the temporary paper file to conserve storage and maintain efficiency.

This multi-step automated pipeline streamlines the data collection process, balancing automation with minimal manual intervention to ensure optimal data integrity. During the data collection process, we face several challenges:\\
\textbf{i. Reading PDFs and Extracting References:} Parsing PDFs to extract reference numbers and cited sentences presents a significant challenge. Variations in textual formatting and encoding across different documents make the task more complex. We use Python-based tools~\cite{pedregosa2011scikit} to automate and assist with this process.\\
\textbf{ii. Converting Table References:} In some cases, references appear in tabular form. Converting these entries into clear sentences or association rules requires careful interpretation and manual verification.\\
\textbf{iii. Formatting Extracted Points:} We format the extracted sentences to focus solely on the relevant reference while removing extraneous words or phrases. This step ensures that the extracted text remains concise and that reference numbers are preserved as they appear in the original document.\\
 \ignore{   
    \begin{itemize}
        \item \textbf{Input:} ``There are tools for automatically reproducing in-field failures from various sources, including core dumps [73, 81], function call sequences [42], call stack [74], and runtime logs [77, 78].''
        \item \textbf{Output:} ``There are tools for automatically reproducing in-field failures from various sources, including runtime logs [77, 78].''
    \end{itemize}.}
\textbf{iv. Handling CAPTCHA Challenges:} Some sources require CAPTCHA verification, which demands human intervention to continue the data collection process.

\begin{table}[hbt!]
\centering
\caption{Sample Collected Data}
\renewcommand{\arraystretch}{1}
\setlength{\tabcolsep}{5pt}
\scalebox{0.85}{
\begin{tabular}{|p{2.2cm}|p{2.2cm}|p{2.3cm}|p{2.2cm}|}
\hline
\textbf{SCMINER} & \textbf{SIMRACER} & \textbf{DESCRY} & \textbf{RRF} \\
\hline

Techniques like PCA-based anomaly detection [174,164] and frequent pattern mining [174] 
are commonly used in system call sequence analysis for intrusion detection, malware detection, 
and fault localization [77], with tools like SCMiner [164] and data mining approaches 
[26] leveraging system and audit logs to identify application-specific faults or attacks [52]. 
&

Details of shared resource and event modeling [21,38], tool comparisons [21,29,38], active testing
[31,38,57], concurrency bug detection [18,26,35,46,50,57], SimRacer’s testing capabilities [18,21,38,57,149], and related research on process-level races [10,18,21,38,41,57,
152] indicate an extensive focus on system-level concurrency issues [18,24,56,60,95]. 
&

Various tools and methods for reproducing in-field software failures leverage data sources like core dumps [8,49,56,73,83,155], runtime logs [43,51,53,54,77,89], symbolic execution [25,217,208,151], partial-order reduction [76,118], concurrency failure analysis [17,18,32,142], static and dynamic analysis [14,24,217], regression testing [85,151], and interleaving exploration [8,17,153,168,169]. %These approaches facilitate bug reproduction without runtime tracing [151,154] and prioritize interleavings 
[32]. 
&

Several tools have been developed for debugging and identifying concurrency and race condition issues, including RacePro [80], SimRacer [117,153], and frameworks like RRF [55], with related case studies and reports cited in [49-55,63-65,117]. 
\\
\hline
\end{tabular}}
\label{tab:sampledata}
\end{table}

Table \ref{tab:sampledata} presents sample data for four research papers. For each paper, we include cited sentences from different sources, separated by commas in the table.

    %\item \textbf{Parsing PDFs:} PDF parsing was challenging due to variations in textual formatting and encodings across different documents.
    
    %\item \textbf{Missing Paper References:} In some cases, the cited paper did not have the actual paper listed under the references section, which complicated the data collection process further. Despite these challenges, our meticulous approach ensured the collection of a comprehensive dataset essential for our sentiment analysis in software engineering research.
%\end{enumerate}

\subsection{Data Cleaning}\label{AA}
After the data collection phase, we clean the dataset to ensure its quality and reliability for subsequent analysis. This process focuses on removing stopwords \cite{10903889}, numbers, and non-alphabetic characters from the extracted text. Stopwords are common words that frequently occur in a language but carry little semantic value in analytical contexts, examples include ``the,” ``and,” and ``is” etc. Removing these words allows us to concentrate on terms that convey more meaningful information relevant to sentiment analysis.

We also eliminate numbers and non-alphabetic characters from the dataset. Although numbers can be significant in certain contexts, they rarely contribute to the sentiment expressed in textual data and may introduce noise. Similarly, punctuation marks, symbols, and other special characters are removed to maintain consistency and readability.

Using NLTK and regular expressions, we refine the dataset by filtering out these irrelevant elements. This cleaning process enhances the dataset’s overall quality, reduces noise, and ensures that the remaining text is more representative and suitable for accurate sentiment analysis.

\subsection{Data Transformation and Modeling}\label{AA}
After completing the data cleaning phase, we proceed to the analysis stage using unsupervised machine learning technique k-means clustering \cite{ahmed2020k}, which distinguishes our approach from prior research in this field. We implement these methods using Scikit-Learn \cite{pedregosa2011scikit}, a widely used machine learning library in Python. For feature extraction, we employ BERT sentence embeddings via Hugging Face \cite{devlin2019bert}, a state-of-the-art natural language processing framework capable of capturing nuanced contextual and semantic information from text. These embeddings provide rich representations that are essential for effective sentiment classification.

We then train the K-Means clustering algorithm \cite{hartigan1979algorithm} to group the data into three sentiment categories: Positive, Neutral, and Negative. To assess the performance of our clustering, we calculate the Silhouette Score \cite{ROUSSEEUW198753} , which measures cluster cohesion and separation. The score ranges from -1 to 1, with values closer to 1 indicating well-defined and meaningful clusters, while values near -1 suggest poor clustering quality. Finally, we apply t-SNE \cite{maaten2008visualizing} for dimensionality reduction, transforming the high-dimensional embeddings into a two-dimensional space to enable intuitive and visually interpretable cluster visualization.

\subsection{Text Summarization}
We advance our analysis by applying summarization techniques to the extracted data points, categorizing them as positive or negative. Using natural language processing, we extract sentiment from each entry while retaining citation numbers to preserve reference integrity.

To strengthen the analysis, we compare these data-point summaries with full-paper summaries, also categorized by sentiment, generated using Gemini 1.5 \cite{gemini} via LangChain \cite{LangChain} and prompt engineering \cite{Sahoo2024Survey}. By contrasting sentiment outcomes between data points and full papers, we evaluate consistency and accuracy across different aggregation levels.

This multi-tiered approach enables cross-validation, revealing discrepancies or alignments in sentiment classification. It also demonstrates how summarization depth affects sentiment fidelity, highlighting the strengths and limitations of LLMs. Overall, the analysis emphasizes the importance of method selection and clarifies the relationship between summarization granularity and sentiment accuracy.

\section{Experimental Evaluation}
To evaluate \Name{}, we investigate the following research questions:\\
\textbf{RQ1:} How effective is the automated data-collection pipeline used in this study?\\
\textbf{RQ2:} How well does the K-means clustering algorithm perform on the extracted citation dataset?\\
\textbf{RQ3:} How strong is the quality of the summaries generated by generative AI in capturing sentiment specific insights?

\subsection{Evaluation Metrics}
\subsubsection{Unsupervised Sentiment Analysis} To evaluate the performance of our sentiment analysis using the K-Means model, we use the \textit{Silhouette Score} \cite{ROUSSEEUW198753}. The Silhouette Score ranges from -1 to 1, where higher values indicate that a data point fits well within its own cluster, while lower values suggest that it poorly fits surrounding clusters. A score near zero indicates that the data point overlaps between clusters.

The score measures how similar a data point is to its own cluster compared with other clusters. The silhouette coefficient is defined as:

\[s(i) = (b(i) - a(i)) / max(a(i), b(i))\]

Here s(i) provides the score for each data point i, a(i) provides the mean separation among data point i and all other points in the identical cluster, and b(i) provides the mean separation between data point i and all points in the closest cluster that i is not a part of.

\ignore{
\subsubsection{Text Summarization:}

\subsubsection{Prompt Engineering:}\label{AA}

An example would be as follows. The answer section is left blank for the LLM to respond.

 \begin{itemize}
        \item \textbf{Context:} providing the complete paper or extracted citations. 
        \item \textbf{Summary:} providing the summary. 
        \item \textbf{Question:} Evaluate the quality of the following summary based on its relevance, coherence, and informativeness compared to the original text. Provide a score for each criterion on a scale of 1 to 10, and include a brief explanation for your ratings. Then, mention any key details from the original text that are missing or misrepresented in the summary. 
        \item \textbf{Answer:}
        
\end{itemize}
}
\subsection{Summary Comparison:} To measure whether the summary preserves the meaning of the original text, we calculate the \textit{Semantic Similarity Score}~\cite{simantic}. This score directly evaluates how well the generated summary retains the source content. Using advanced language model embeddings, such as BERT or Sentence-BERT, we convert both the original text and the summary into dense vector representations. We then compute the cosine similarity between these vectors, which quantifies the semantic overlap and provides a numerical score reflecting how faithfully the summary conveys the core ideas and context of the source material. Combined, these methods offer a rigorous framework for evaluating summarization models, balancing qualitative insights with quantitative metrics to ensure meaningful and contextually accurate outputs.

\ignore{
The similarity metrics that are used are cosine similarity and are calculated as:

\[
%\text{cosine similarity} = S_C(A, B) := 
\cos(\theta)
= \frac{\mathbf{A} \cdot \mathbf{B}}{\|\mathbf{A}\| \|\mathbf{B}\|}
= \frac{\sum_{i=1}^{n} A_i B_i}{
\sqrt{\sum_{i=1}^{n} A_i^2} \,
\sqrt{\sum_{i=1}^{n} B_i^2}},
\]
}

\section{Results and Analysis}

\subsection{RQ1: Effectiveness of automated Data-Collection}
% \reviewtwo{Clarify experimental scope: Clearly explain why five papers could not be processed and how this affects generalizability.
% }

To evaluate the effectiveness of our automated data-collection process, we assess its ability to extract citation statements along with their corresponding reference numbers from nine research papers. The system successfully extracts citation data for four papers: SCMiner \cite{8952396}, SimRacer \cite{10.1145/2483760.2483771}, DESCRY \cite{yu2017descry}, and RRF ~\cite{7774517}. All of these four papers provide consistent formatting and few barriers to text parsing, enabling smooth and accurate extraction.

The process encounters challenges with the remaining five papers due to several obstacles:
i) Captcha Verification: Automated scripts cannot bypass captcha requirements on certain platforms,
ii) Text Parsing Errors: Irregular formatting and complex reference structures cause parsing failures, and iii) Timeout Issues: Data retrieval exceeds time limits, especially for large or access-restricted documents.

As a result, the overall success rate is 44\%, revealing the limitations of the current automated approach when dealing with diverse document formats and access constraints. Despite these challenges, the successfully extracted cases offer useful insights for refining the methodology. They also underscore the need to address access restrictions and improve parsing algorithms to enhance the system’s robustness and scalability. Consequently, we maintain data collection as a semi-automated process and introduce manual intervention for cases where the automated system fails.

\subsection{RQ2: Performance Evaluation of K-means:}

We  present  the  K-mean  performance  on  two  and  three  clusters  which  was  visualized  using t-SNE~\cite{maaten2008visualizing} and evaluated using Silhouette score~\cite{ROUSSEEUW198753}. 

\begin{figure}[htbp]
\centerline{\includegraphics[scale= 0.36]{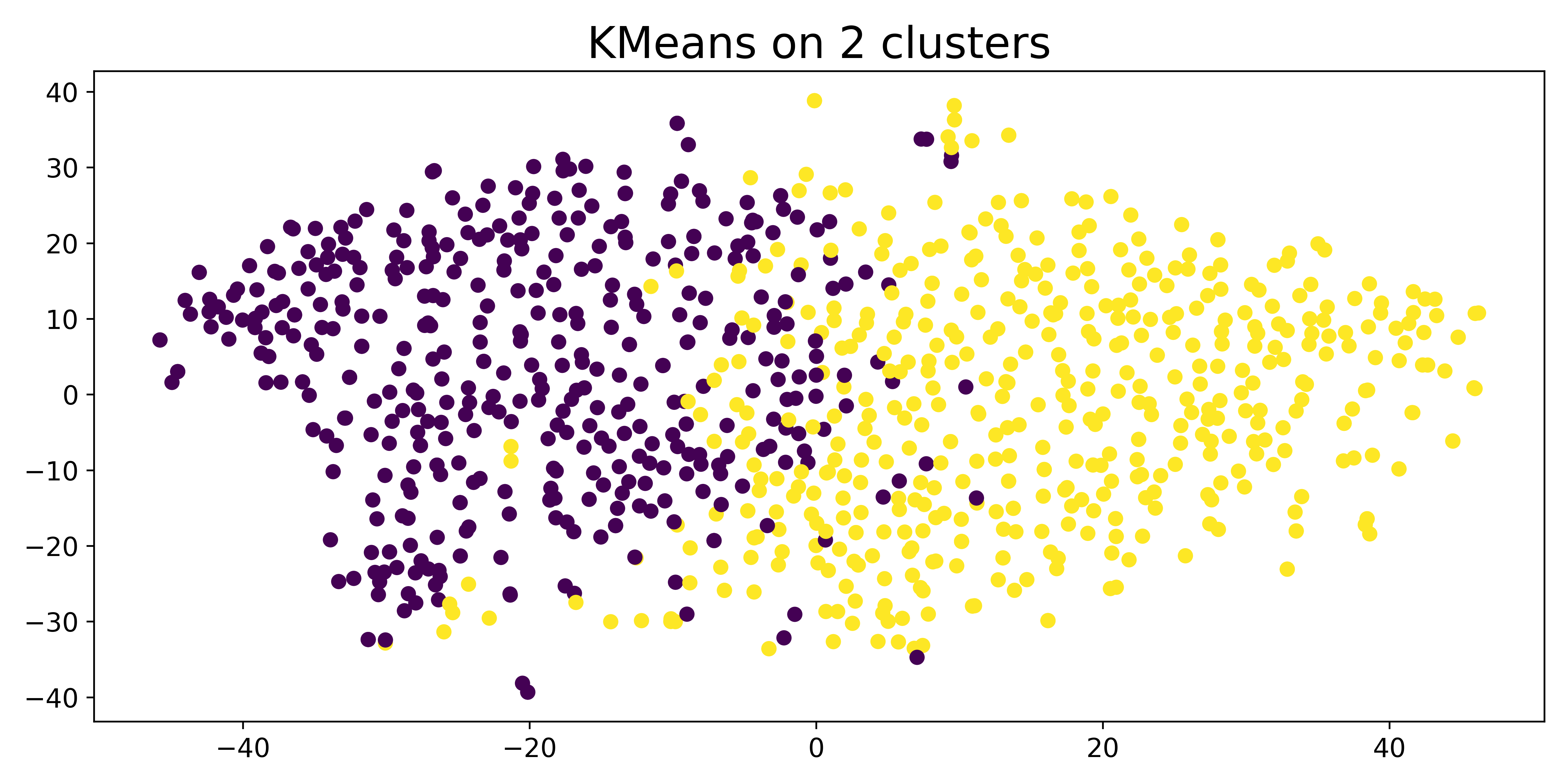}}
\caption{Performance on Two Clusters}
\label{fig:k2}
\end{figure}

Figure \ref{fig:k2} presents the results of applying a K-Means clustering algorithm to our dataset to form two clusters, shown in yellow and dark blue. The data points span approximately –50 to 50 on the x-axis and –40 to 40 on the y-axis. The algorithm groups points based on their feature similarities, illustrating how clustering can reveal underlying patterns in the data. However, the Silhouette Score of 0.1384 indicates weak separation between clusters, suggesting substantial overlap or limited internal cohesion within each cluster.
%This score implies that the effectiveness of the clustering could be improved \commenthm{could be improved-1}, possibly by adjusting the number of clusters or fine-tuning the algorithm's parameters. 

While the K-Means algorithm successfully groups the data into two clusters, the low Silhouette Score underscores the need for further optimization to achieve clearer and more distinct separation. Improving clustering techniques remains essential in machine learning and data analysis, as enhanced cluster quality enables more reliable pattern recognition and data categorization.

\begin{figure}[htbp]
\centerline{\includegraphics[scale=0.36]{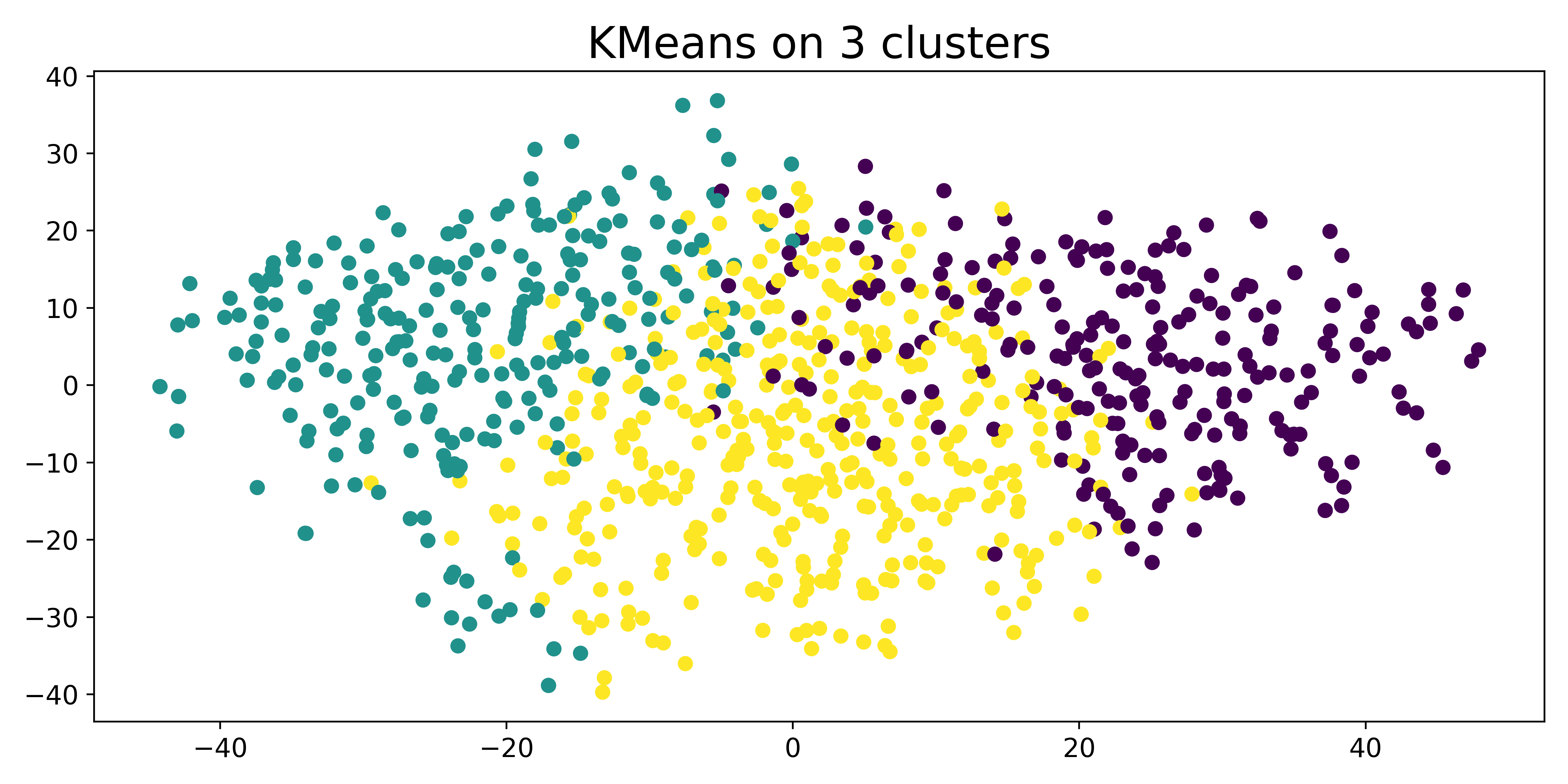}}
\caption{Performance on Three Clusters}
\label{fig:k3}
\end{figure}

Figure~\ref{fig:k3} shows the results of applying the K-Means algorithm to generate three clusters, displayed in teal, purple, and yellow. The scatter plot presents data points spread across a two-dimensional space with x and y values ranging from roughly –40 to 40. Although the algorithm partitions the data into three groups, substantial overlap remains, indicating weak separation among the clusters.

The Silhouette Score of 0.1086 further reflects the low clustering quality. This score suggests that the clusters are not well defined and that the data points within each group are not clearly distinguishable from those in other clusters.

%Such a score indicates that there could be improvements made to the clustering approach \commenthm{could be improved-3}, such as adjusting the number of clusters or refining the algorithm's parameters, to achieve more distinct and meaningful groupings in the data. Understanding and enhancing these techniques are essential in data analysis and machine learning for better pattern recognition and data organization.

\begin{table}[hbt!]
\caption{Score Comparison}
\centering
\begin{tabular}{|l|l|l|l|l|}
\hline
\textbf{} & \textbf{2 Clusters} & \textbf{3 Clusters} \\
\hline
\textbf{Silhouette  Score} & 0.138 & 0.108 \\
\hline
\end{tabular}
\label{Tab:shil}
\end{table}

Table \ref{Tab:shil} shows that K-Means with two clusters, positive and negative, outperforms K-Means with three clusters (Positive, Neutral, and Negative). However, due to ambiguities in citation sentence structure and formatting, the highest Silhouette Score achieved is only 0.138 for the two-cluster model. Despite low scores, K-means remains appropriate for this task because it provides a systematic method to partition citation statements into meaningful sentiment groups (positive/negative). Therefore, we select the two-cluster K-Means model and apply it to the four papers.

\subsection{RQ3: Evaluation of the Summaries}
%\commenttsz{I am not sure about this section. What did you do here? Can you please tell me?}
To evaluate the quality and accuracy of the summaries we conduct two types of experiments, i) Empirical study, and ii) Manual study.

To perform this evaluation we use Gemini Flash 1.5 to generate summaries. First, we instruct Gemini Flash 1.5 to generate summary of the strength of a paper by only providing the positive feedback provided by \Name{}. Table \ref{tab:positive} presents the summaries generated by using \Name's positive feedback. Similar workflow is done for negative feedback. 

%We also instruct GPT 4 \commenthm{mention gpt 4 in Text summerization section?} \commenthm{cite? first mention of gpt in current version} to generate strentghs  and limitations of the paper by providing the full research paper. We then comapre these results manually. 

\subsubsection{Empirical Study}
%We provide Gemini Flash 1.5 with the positive results generated by \Name{} for each paper and instruct it to produce a summary highlighting the strengths of the paper based on this feedback. Additionally, we supply the full research paper and direct the LLM to summarize the strengths of the work from the complete document.

\begin{table}[hbt!]
\caption{Comparison of Summaries }%\commenthm{combine with table 2?}}
\centering
\begin{tabular}{|l|l|l|l|l|}
\hline
\textbf{} & \textbf{SCMINER} & \textbf{SIMRACER} & \textbf{DESCRY}  & \textbf{RRF} \\
\hline
\textbf{Sim.  Score} & 0.58 & 0.68 & 0.60 & 0.70\\
\hline
\end{tabular}
\label{tab:simscore}
\end{table}

Table \ref{tab:simscore} compares the similarity scores of two summaries for four research papers: SCMINER, SIMRACER, DESCRY, and RRF. The similarity score measures how closely the two approaches align in summarizing each paper. Overall, the scores show varying degrees of alignment, with an average similarity of 0.64 across the summaries.

\begin{table*}[hbt]
\centering
\caption{Strengths of Research Papers Based on \Name{}’s Positive Feedback}
\renewcommand{\arraystretch}{1.3} % increases row height
\small % increases font size
\scalebox{0.7}{
\begin{tabular}{|p{3.3cm}|p{2cm}|p{1.8cm}|p{2.5cm}|p{2.3cm}|p{2.3cm}|p{1.8cm}|p{2.3cm}|p{2.3cm}|}
\hline
\textbf{SCMINER} & \textbf{SIMRACER} & \textbf{DESCRY} & \textbf{RRF} & \textbf{ReDPro} & \textbf{DAMCBMmE} & \textbf{PDPRDS} & \textbf{FCESMCPR} & \textbf{DRDDR} \\
\hline

SCMINER automates the localization of system-level concurrency faults and was empirically evaluated on 19 real-world Linux applications. It does not require multiple failed runs and uses Principal Component Analysis (PCA) for anomaly detection.

& SimRacer automates testing for process-level race conditions by controlling process scheduling to reproduce real-world races.

& DESCRY automates the reproduction of system-level concurrency failures and supports large-scale Linux applications.

& RRF combines static program analysis and dynamic event reporting, effectively reproducing and debugging process-level races caused by timing and interleaving issues.

& ReDPro provides automated detection and regeneration of process-level concurrency failures using the open-source PIN binary instrumentation tool.

& The proposed dynamic analysis method detects concurrency bugs in multi-process and multi-thread environments by intercepting execution with optimized data collection.

& RACEPRO’s OS-level in-vivo model checking detects process races in live systems with minimal disruption.

& RACEPRO also provides in-vivo model checking that detects races in deployed systems without source code, addressing both read-write and write-write races efficiently.

& The study on cross-race effect (CRE) uses a dual-process approach to analyze recollection and familiarity differences in own-race vs. other-race face recognition.
\\
\hline
\end{tabular}}
\label{tab:positive}
\end{table*}

\subsubsection{Manual Analysis}
We also perform a manual analysis to evaluate the summaries generated by LLMs. To do this, we assess three metrics: (i) conceptual correctness, (ii) factual correctness, and (iii) informativeness. We assign a score from 1 to 5 for conceptual correctness based on how well the summaries capture the central themes of the papers. For factual correctness, we score the summaries from 1 to 5 according to their logical flow and accuracy. Informativeness measures how comprehensively the summaries convey the research details, also rated on a scale from 1 to 5.
%These metrics demonstrate Gemini Flash's ability to produce concise and structured summaries.

\ignore{
\begin{figure}[htbp]
\centerline{\includegraphics[width=0.5\textwidth]{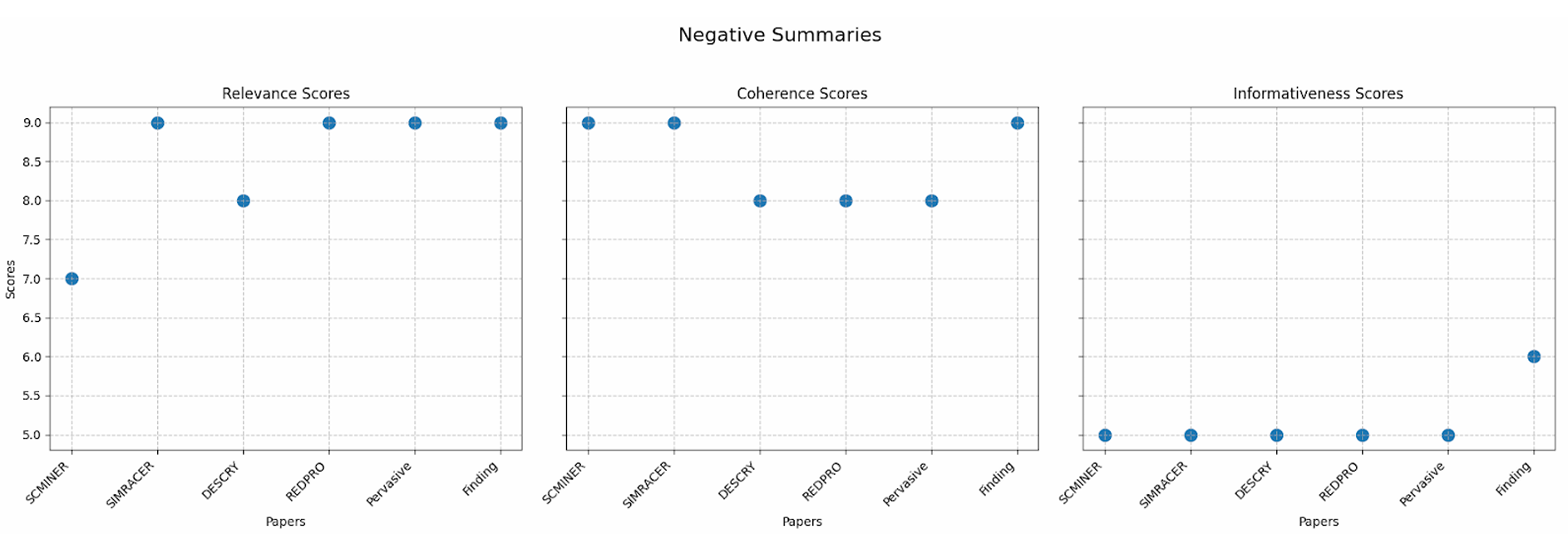}}
\caption{Negative Summaries}
\label{fig}
\end{figure}

\begin{figure}[htbp]
\centerline{\includegraphics[width=0.5\textwidth]{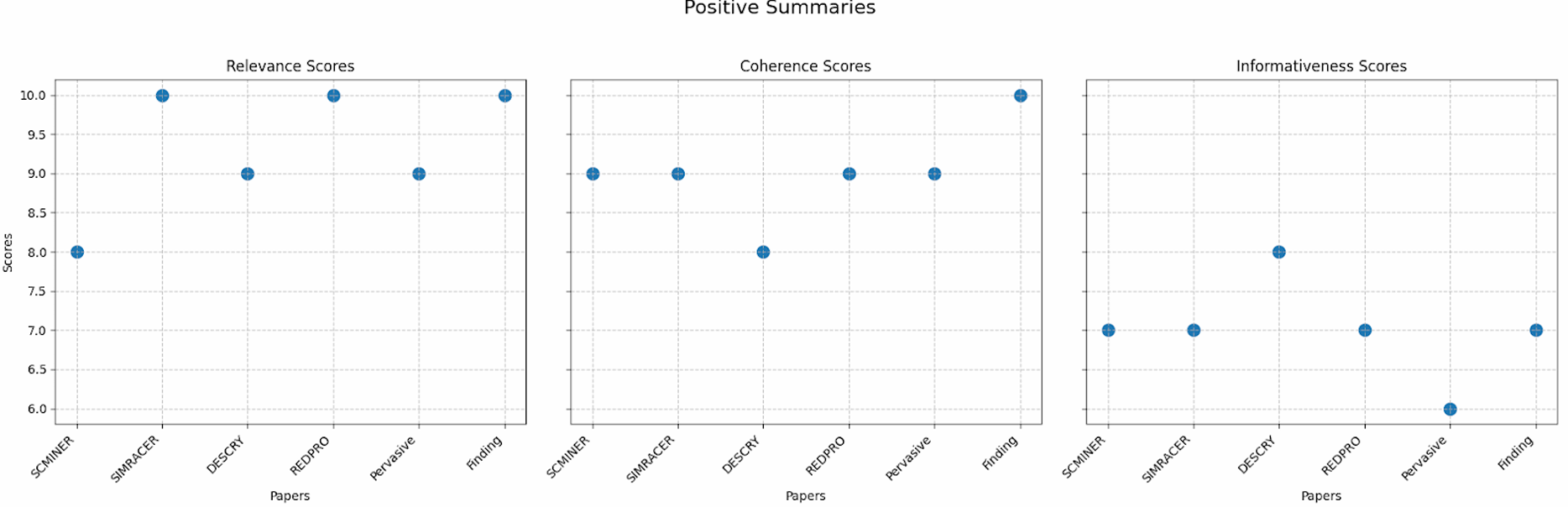}}
\caption{Positive Summaries \commenthm{text not visible. can we add the graphs as pdf and may be in two column?}}
\label{fig}
\end{figure}
}
\begin{figure}[htbp]
\centerline{\includegraphics[scale=0.65]{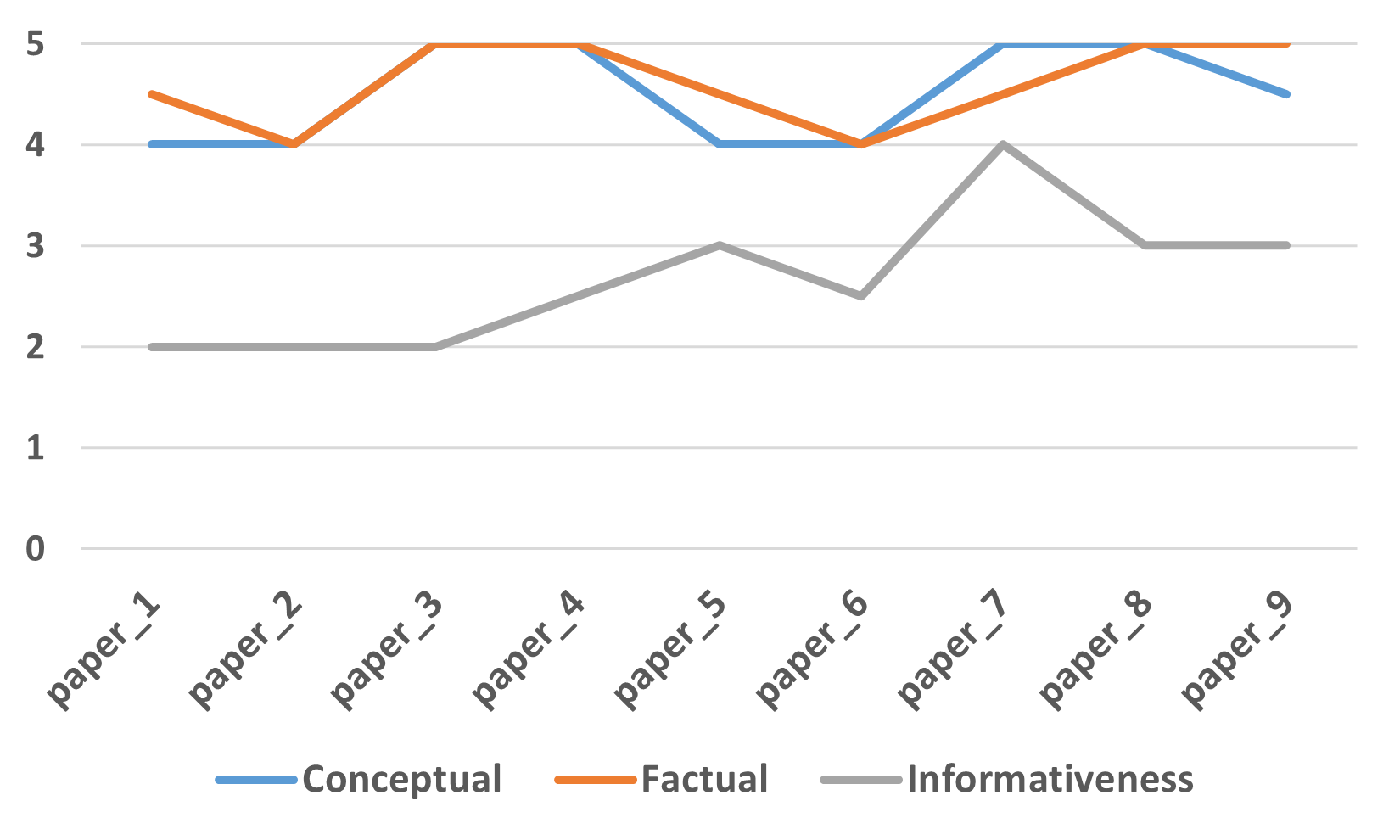}}
\caption{Manual Analysis on Summaries}
\label{fig:Manual}
\end{figure}

% \reviewtwo{Enhance figure readability: Figures 2–4 should include clearer legends, axis labels, and short captions summarizing key insights.
% }

Figure \ref{fig:Manual} presents our manual evaluation scores (1 to 5) for nine research papers. Positive summaries generated by Gemini Flash 1.5 achieve high conceptual and factual scores, with some papers receiving perfect ratings, demonstrating the model’s ability to capture core ideas and maintain logical structure. Informativeness scores are slightly lower, indicating limitations in depth and detail. Negative summaries show reduced conceptual correctness and informativeness, although factual accuracy remains stable. These results provide a clear assessment of summary performance across the key metrics.

%Gemini Flash 1.5-generated positive summaries achieve high relevance with some papers scoring near perfect, reflecting the model's ability to capture core ideas and maintain logical structure. However, informativeness and coherence scores are slightly lower, showing limitations in depth and comprehensiveness. These results provide a clear picture of the performance \commenthm{good or bad?} of the summaries across key metrics.

%\subsubsection{Validating the Summaries from citations using GPT-4o}

\ignore{
\begin{figure}[htbp]
\centerline{\includegraphics[width=0.5\textwidth]{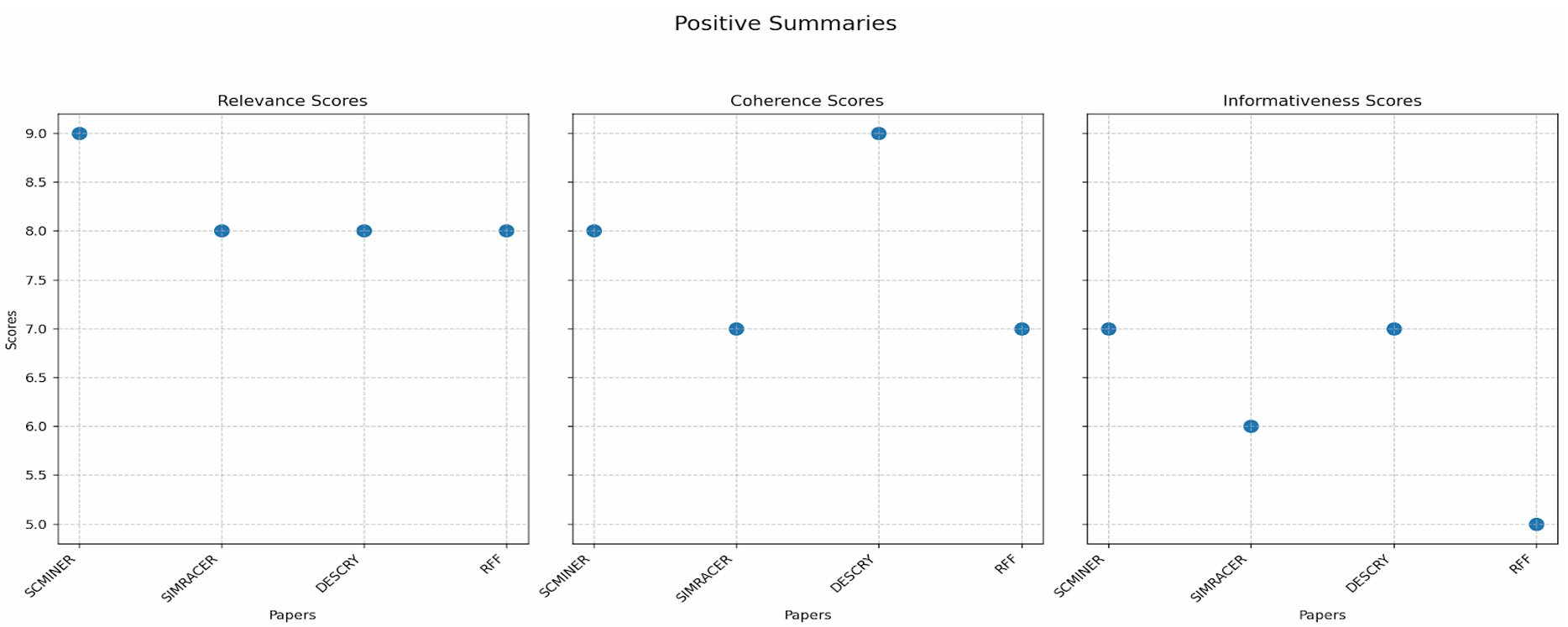}}
\caption{Positive Summaries}
\label{fig}
\end{figure}}

%\commenthm{start new section "Related Work"?}
\section{Related Work}
Citation sentiment analysis has gained attention as a method to understand how authors perceive prior work. Early studies in the biomedical domain, such as Automated Citation Sentiment Analysis: What Can We Learn from Biomedical Researchers (2013, 2014), classify citations as positive, neutral, or negative to improve citation-based metrics, highlighting the challenges of technical and precise domains \cite{yu2013automated}.

Subsequent approaches incorporate contextual and linguistic information to enhance sentiment detection. For example, Context-Enhanced Citation Sentiment Detection (2012) and Aspect-Based Citation Sentiment Analysis Using Linguistic Patterns (2019) leverage surrounding text and specific citation aspects for more nuanced insights \cite{athar2012context, ikram2019aspect}. Meta-analyses extend this by linking sentiment to citation intent, considering role and function alongside polarity \cite{kunnath2021meta, budi2023understanding}.

Recent studies employ AI and deep learning to improve scalability and accuracy. Neural network–based models detect sentiment across diverse datasets \cite{zhou2022research, karim2022comprehension, muppidi2020approach}, while integration with text mining methods, such as TF-IDF and ML classifiers, further enhances classification. Domain-specific work, including clinical trial analyses, evaluates citation sentiment in context to assess trustworthiness and impact \cite{xu2015citation}. Techniques for identifying influential citations leverage sentiment-rich textual features and resampling methods like SMOTE to improve model robustness \cite{nazir2022important, umer2021scientific}.

Large-scale approaches utilize citation databases combined with linguistic or deep learning methods for systematic analysis \cite{parthasarathy2014sentiment}. Additionally, contextualized summarization studies integrate citation sentiment to improve literature comprehension. Tandon and Jain (2012) categorize sentiments in citation contexts to infer relevance and aid summarization \cite{tandon2012citation}, while Syed et al. (2023) propose citance-contextualized summarization, incorporating sentiment, intent, and citation function to generate more informative summaries \cite{syed2023citance}.

These studies collectively establish a foundation for combining sentiment analysis with citation-aware summarization, improving understanding, evaluation, and synthesis of scholarly work. In our work \Name{}, we analyze both positive and negative citation feedback and generate summaries based on this information. Our goal is to determine whether citation-based summaries provide more concrete and useful insights than traditional summaries, particularly regarding the strengths and limitations of a paper.

\section{Discussion and Limitations}

% \reviewtwo{Expand limitations section: Add details on citation-context ambiguity, reference extraction noise, and domain bias toward software engineering papers.
% }

While the \Name{}  framework introduces a novel approach to understanding scholarly impact, several aspects offer opportunities for further enhancement.

The semi-automated citation extraction process, for instance, faces technical challenges such as captcha verification, text parsing errors, and occasional timeouts. Additionally, challenges in determining optimal citation context boundaries and PDF parsing noise including formatting artifacts and incomplete bibliographic data may affect analysis accuracy. These challenges currently limit the extraction success rate, highlighting opportunities to refine data acquisition methods and improve robustness across diverse document formats.

\Name{} leverages K-Means clustering, which performs well in many scenarios but can be sensitive to the initial placement of centroids  \cite{arthur2006k} and less effective on imbalanced datasets \cite{ahmed2020k}. Future work could explore alternative or adaptive clustering techniques such as DBSCAN \cite{ester1996density}, HDBSCAN \cite{campello2015hierarchical}, Spectral Clustering \cite{ng2001spectral} etc. to enhance grouping accuracy and better capture the nuances of citation sentiment.

Although transformer-based LLMs provide powerful summarization and sentiment classification capabilities, their internal decision-making can appear opaque. This presents an opportunity to incorporate explainability techniques to better understand which features and linguistic patterns influence the model’s outputs.

The study focuses on software engineering literature and a limited sample of papers, which provides a strong proof-of-concept while suggesting avenues for broader evaluation. Domain bias toward software engineering papers may limit generalizability, as citation patterns, terminology, and sentiment expressions may vary significantly across academic disciplines. Extending \Name{} to other domains and diverse datasets can test its adaptability to different citation behaviors and textual characteristics.

%Finally, while the current framework operates on English-language citations, applying SECite to multilingual scientific literature remains an open direction. Expanding language support would increase the framework’s applicability in global research contexts.

%Overall, these limitations highlight areas for growth and refinement, while demonstrating the framework’s potential to support meaningful insights into citation sentiment and scholarly impact.

%Finally, as citation intent and sentiment are subjective in nature, the lack of extensive user studies or expert validation of SECite limits its practical usefulness and real-world application.

\section{Conclusion and Future Work}

This study investigates the positive and negative aspects of research papers from two perspectives: external citations and internal content. We develop a semi-automated approach to extract and analyze citations, categorizing them into positive and negative sentiments using K-Means clustering. Simultaneously, we leverage large language models (LLMs) to generate sentiment summaries directly from the full text of the papers.

Applying this methodology to nine research papers reveals how the scholarly community perceives these works and how the papers present themselves. Comparing citation-based and full-text perspectives highlights both alignments and divergences, offering a more holistic view of each paper’s impact.

Future work will focus on enhancing automation and accuracy across multiple stages of the process. Key priorities include improving the reliability of citation discovery, parsing, and extraction by addressing challenges such as captcha handling, text parsing errors, and inconsistent formatting. For sentiment classification, refining the K-Means algorithm or exploring advanced alternatives, such as zero-shot prompt engineering with LLMs, could improve the grouping of citations without requiring labeled data.

Additionally, evaluation metrics for summaries can be further developed to better capture coherence, completeness, and relevance for both citation-based and full-text summaries. These advancements aim to make the framework more efficient, scalable, and insightful, supporting broader applications in understanding scholarly impact and research influence.

\section{Acknowledgment}
This work was supported in part by NSF grants CCF-2348277 and CCF-2518445.

% \bibliographystyle{IEEEtran}
% \bibliography{bibfile}


% \vspace{12pt}
% \color{red}
% IEEE conference templates contain guidance text for composing and formatting conference papers. Please ensure that all template text is removed from your conference paper prior to submission to the conference. Failure to remove the template text from your paper may result in your paper not being published.

\end{document}